\documentclass[12pt]{article}
\usepackage{graphicx}
\usepackage[top=3cm, bottom=2cm, left=3cm, right=3cm]{geometry}
\title{Fine structure in the cosmic ray spectrum: further analysis and the next step.}
\author{A. D. Erlykin $^{1,2}$ and A. W. Wolfendale $^{2}$\\
$(1)$ P N Lebedev Physical Institute, Moscow, Russia\\
$(2)$ Department of Physics, Durham University, Durham, UK}
\begin{document}
\maketitle

\begin{abstract}
An analysis is made of the fine structure in the cosmic ray energy spectrum: new facets
 of present observations and their interpretation and the next step. It is 
argued that less than about 10\% of the intensity of the helium `peak' at the knee at 
$\simeq 5PeV$ is due to just a few sources (SNR) other than the single source. The 
apparent concavity in the rigidity spectra of protons and helium nuclei which have 
maximum curvature at about 200 GV is confirmed by a joint analysis of the PAMELA, 
CREAM and ATIC experiments. The spectra of heavier nuclei also show remarkable 
structure in the form of `ankles' at several hundred GeV/nucleon. Possible mechanisms 
are discussed. The search for `pulsar peaks' has not yet proved successful.
\end{abstract}

\section{Introduction}
The search for the origin of Cosmic Rays (CR) is a continuing one. Although many
incline to the view that supernova remnants (SNR) are responsible, it is also possible
that pulsars contribute and other possibilities include
distributed acceleration. To distinguish between them is not a trivial task but is
attempted here, by way of studies of the detailed shape, or `fine structure', of the CR
 spectrum.

Our claim that a `single source' is largely responsible for
the characteristic knee in the spectrum (Erlykin and Wolfendale, 1997, 2001) has 
received support from later, more precise, measurements. Very recently (Erlykin and
Wolfendale, 2011a,b), we have put forward the case for further fine structure in the 
energy spectrum in the knee region
which appears to be due to the main nuclear `groups': P, He, CNO and Fe, but this
awaits confirmation. What is apparent, however, is that there should be such fine
structure in the spectrum and that this structure should give strong clues as to the
origin of CR.

It is appreciated that there is the danger of 'over-interpreting the data' but, in our 
view, this is the only way that the subject will advance. For many years, new 
measurements merely confirmed that 'there is a knee in the CR energy spectrum' without 
pushing the interpretation forward. Our own hypothesis met with considerable scepticism
 initially, and, indeed, there are still some doubters but we consider that the new 
data are sufficiently accurate for 'the next step' to be taken. Only further, more 
refined data will allow a decision to be made as to whether or not the present claims 
are valid.

We start by exploring the properties of SNR and pulsars from the CR standpoint and go
on to examine the present status of the search for SNR-related structure by way of
studying the world's data on the primary spectra and the relationship of various
aspects to our single source model. Particular attention is given to the contribution
to the single source peak from the second strongest source and further 'structure', in
 the form of curvature in the energy spectra of the various components. The situation
 in the region of hundreds of GeV/nucleon is considered. Returning to pulsars, the 
result of a search for 'pulsar-peaks' is described.

\section{Comparison of SNR and Pulsars.}
\subsection{Supernova Remnants.}
Type II SN, which are generally regarded as the progenitors of CR, have a Galactic
frequency of $\sim 10^{-2}$y$^{-1}$ and a typical total energy $\sim 10^{51}$erg for 
each one. Most models (eg Berezhko et al. 1996) yield $\sim 10^{50}$erg in CR up to a 
maximum rigidity of $\sim 3$PV. The differential energy spectrum on injection is of the
 form $E^{-\gamma}$ where $\gamma \sim 2$ (or a little smaller). At higher rigidities, 
not of concern here, models involving magnetic field compression via cosmic ray 
pressure can achieve much higher energies (eg Bell, 2004).

\subsection{Pulsars.}
 An alternative origin of CR above the knee is by way of pulsars which, although being 
thought to be barely significant below 1-10 PeV, may well predominate as CR sources at 
higher energies (~eg Bednarek and Bartosik, 2005~). If there are small contributions 
below 1 PeV, however, they might give rise to small peaks (~see \S4.2 later~). 
Unlike SNR there are no `typical' pulsars in that their initial
periods differ, depending as they do on the rotation rate of the progenitor star. With
the conventional value for the moment of inertia of $10^{45}g cm^2$, the rotational
energy $(\frac{1}{2}$I$\Omega^2)$ is $2\times 10^{52}P_{ms}^{-2}$erg, where $P_{ms}$ is
 the period in ms. SN and pulsars thus have similar total energies only if the pulsar
has an initial period less than about 4.5ms. The fraction of energy going into CR is
not clear but may be in the range (0.1-1)$\%$ (Bhadra, 2005).

The maximum rigidity is given by $R_{max} = 6.6\cdot10^{18}B_{12} P_{ms}^{-2}$ V
(Giller and Lipski, 2002) where $B_{12}$ is the effective magnetic field, in $10^{12}$
Gauss and, as before, $P_{ms}$ is the period in ms. Insofar as there is, presumably, a
 significant fraction of pulsars with birth periods less than 10ms there appears to be
no problem in achieving the necessary PeV energies. Indeed, the case has been made for
the rest of the CR energy range from PeV to EeV being due to fast pulsars, as already
remarked.

The differential energy spectrum of all CR emitted by the pulsar during its age will
 probably be of the form $E^{-1}$ due to there being a delta function in energy at a
particular instant and the energy falling with time as the pulsar loses rotational
energy.

The result of the above is that pulsars cannot easily be involved as the source of 
particles of energy below the spectral knee at $\simeq$3 PeV; however, they cannot be 
ruled out as being responsible for some of the fine structure there.

\section{The present status of the fine structure of the energy spectrum.}
Our first publication claiming evidence for a single source (Erlykin and Wolfendale,
1997) used measurements from 9 EAS arrays. Four of the arrays included data above logE
 $\sim$ 7.5 with E in GeV. A summary of the sharpness S vs log$(N_e/N_e^{knee})$ from
that work is
given in Figure 1a. The sharpness is given by S=$-d^2(logN_e^3I(N_e) / d(logN_e)^2$
where $N_e$ is the shower size and $N_e^{knee}$ is the shower size at the knee
position (the sharpness for the energy spectrum has $N_e$ replaced by E).
Prominent features in the initial plot are represented by A, B, C and D. 'A' was the
concavity first referred to by Kempa et al. (1974). 'B' was the main peak (knee) and
'C' a subsidiary peak; initially, we identified 'B' with CNO and 'C' with Fe but 
latterly,
following direct measurements to higher energies than before, and other indications, 
'B' is identified with He and 'C' with CNO (Erlykin and Wolfendale, 2006). 'D' is an
important minimum identified now as the dip between CNO and Fe. The small peak `E' was
not identified in the 1997 work but it is now realised to be coincident with the Fe 
peak.
\begin{figure}[hptb]
\begin{center}
\includegraphics[height=14cm,width=17cm]{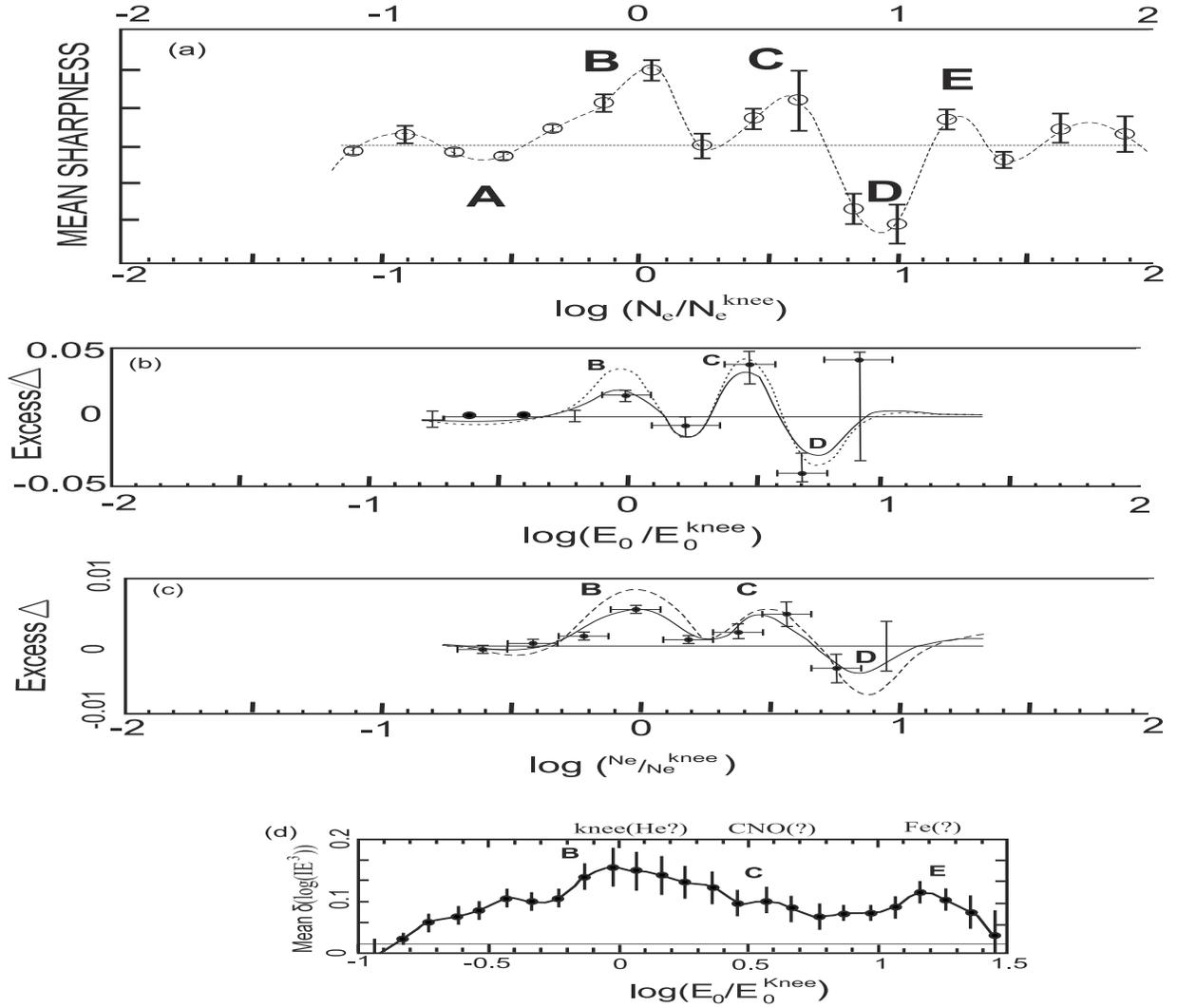}
\end{center}
\caption{\footnotesize Summary of fine structure in the size spectrum (and energy 
spectrum) of cosmic rays from the world's EAS arrays.}

{\footnotesize (a) Spectral sharpness values, S, for size spectra from the initial work
 of Erlykin and Wolfendale (1997). The letters A to D were in the initial paper and 
represented particular features. E is added in view of the most recent work (see (d))
showing a peak identified by us as due to iron. The letters are carried forward
 into the plots given below. Smooth lines in panels (a) and (d) below are cubic 
splines, which pass precisely through the points to guide the eye.

\noindent (b) Fine structure in the Cherenkov light spectra from the work of
Erlykin and Wolfendale (2001). The results are independent of those in (a). The
 excess, $\Delta$(in log(E$^3$I)) is from a 5-point running mean. The dashed
line was the previous prediction, based on the fit to figure 1(a) and the
full-line the prediction from a more comprehensive analysis. The latter fit 
($\chi^2/ndf=0.86$) is better than for the case with no structure ($\chi^2/ndf=4.2$).

\noindent (c) As for (b) but for electron size spectra. There is modest
duplication of data used also in (a). As in the previous panel (b) the full line fit, 
although not good enough ($\chi^2/ndf=1.6$), is certainly better than that for the case
 with no structure ($\chi^2/ndf=15$).

\noindent (d) Fine structure in the primary cosmic ray energy spectrum shown as
 the difference,
$\delta$(in log(E$^3$I)) between the observations and expectation for the
Galactic Diffusion Model, with its slowly steepening spectrum. Emphasis is
placed here on data at energies above those of the knee. The measurements are
from 10 arrays and refer to new data published since those reported in (a), (b)
and (c). (See Erlykin and Wolfendale, 2011a, b). The non-observation of a 'CNO peak'
 (and associated minima between B and C, and D) does not invalidate its presence, as 
discussed in the text; there is clearly an excess intensity to be occupied by some 
nuclei.}

\label{fig:fig1}
\end{figure}

An `update' to the structure problem was given in 2001 (Erlykin and Wolfendale, 2001)
where we analysed data from 16 arrays. In many cases, spectra were given for different
zenith angles; the result was that 40 spectra contributed. The measurements superseded
the 1997 results but only to a limited extent because the later measurements did not 
always cover a much longer period. In that work the difference in intensity from the 
running mean was determined from the Cherenkov data and from the 40 EAS electron size 
spectra  with the results shown in Figures 1(b) and 1(c). We have 
adopted the same nomenclature for the various features, A, B\dots E. 

Finally, and most recently, Figure 1d gives the results from an analysis of 10
more spectra published after 2001 (and studied by Erlykin and Wolfendale, 2011a,b), the
 virtue of these most recent spectra is that they
 extend to higher energies than most of the previous ones and give good statistical
precision for region E (the Iron peak). It will be noted that yet another ordinate is
used, $\delta$(logE$^3$I), the difference of the intensity from that expected from a
`conventional model' - the Galactic Diffusion Model - in which no structure is present,
the datum being a smooth gradually steepening spectrum.

The lesson to be learned from Figure 1 is that the succeeding analyses give consistent
results for the presence of fine structure in the spectra (size spectra and energy
spectra); the difference between features in the various plots is attributed to the
non-linear dependence of shower size ($N_e$) on energy ($E$) and experimental and model
 `errors'. The lack of observation of the peak C(CNO) in the latest data presumably 
arises from both 'statistics' and the fact that the arrays were designed to give higher
 accuracy at high energies (~i.e. regions D and E~). Indeed, very recent observations 
from TUNKA-133 (~Kuzmichev et al., 2011~) and GAMMA (~Martirosov et al., 2011~), not 
shown in Figure 1, give further strong evidence for the Iron peak (~ie 'E' in  
Figure 1a~).

To be consistent with the analyses of Figures 1a to 1c we can examine the $\chi^2$ 
values for Figure 1d, too. As remarked above, the new arrays contributing to Figure 1d 
were designed to go beyond the knee and, accordingly, the 'resolution' of A, B, C and 
D were not as good. Taking this region first, ie, $log(E/E^k)$ from -1 to +1, a smooth 
line can be drawn through the points and the reduced $\chi^2$ value determined. It is
1.0, i.e. a reasonable fit, with no evidence for peaks B or C. However, the datum is 
$\Delta = 0$, corresponding to the conventionally expected value, so that there is 
strong evidence for the excess caused by B plus C (~at the many standard deviations 
level~). The problem is the lack of the usual dip between B and C; bearing in mind the
fact that the 'zero datum'differs between 1d and the rest, the deficit is at the 2 
standard deviation level, only - a not very serious problem.

Turning to the region above $log(E/E^k)=1.0$, there is clear evidence from Figure 1d 
for the peak 'E'. With respect to the datum it is present at the 4.6 standard deviation
level, based on a single point, and about 7 standard deviations if all the points are 
used. In only Figure 1a can a comparison be made; here the significance is at the 2.6 
standard deviation level.

The conclusion is that all the peaks, B, C and E are very significant. Specifically, 
for the single points at the appropriate energies/shower sizes, alone, we have, in 
standard deviations: \\
B ~ 7.6, 4.2, 9.4 and 4.4 \\
C ~ 2.4, 2.6, 3.3 and 3.3 \\
E ~ 2.6 and 4.0 (7) 
\section{The search for further structure.}
\subsection{General remarks.}
`Structure' can be divided into two broad and overlapping categories: sharp
discontinuities (fine structure) and slow trends. Anything other than a simple power
spectrum with energy-independent exponent can be termed `structure'.

The main fine structure is the very well-known knee at $\sim3$PeV, and this will be
examined here in more detail from the standpoint of the contribution from not just the
single source (SNR) but from the very few most recent/nearest sources. These
will have a possible blurring effect on the sharpness of the knee. A related matter for
 SNR is the expected `curvature' of the energy spectrum in the energy region where
direct measurements for various nuclei have been made (up to about $10^5$GeV for
protons and $10^3$GeV/nucleus for iron). Such curvature has been seen in our
calculations (Erlykin and Wolfendale, 2005, see also Figure 5 below) for the SNR model 
in which particles from a random collection of SNR, in space and time, were followed.

Of particular relevance to the pulsar case is the search for sharp discontinuities, 
such as would arise from specific pulsars.

We start with the possibility of more than one source contributing to the knee.

\subsection{Contribution from a `few' nearby sources.}
At any one energy there must be, in principle, contributions from very many sources. By
 dividing the intensity up into contributions from a background formed by the very many
 sources and the `single' source (the `nearest and more recent' - allowing for 
propagation effects) we have, hitherto, ignored those few other sources which are not 
too far away nor too old (or too young) to contribute. The situation is similar to the 
scattering of charged particles passing through an absorber, where there is multiple 
scattering, single scattering and plural scattering; here, we examine the likely 
contribution from `plural' sources. The plural sources have relevance to the sharpness 
of the knee because their own knees will be at somewhat different energies due to  
differences in the values of the parameters which determine $R_{max}$ (SN energy, ISM 
density, magnetic field and degree of field compression, etc.)
\begin{figure}[htb]
\begin{center}
\includegraphics[height=15cm,width=9cm,angle=-90]{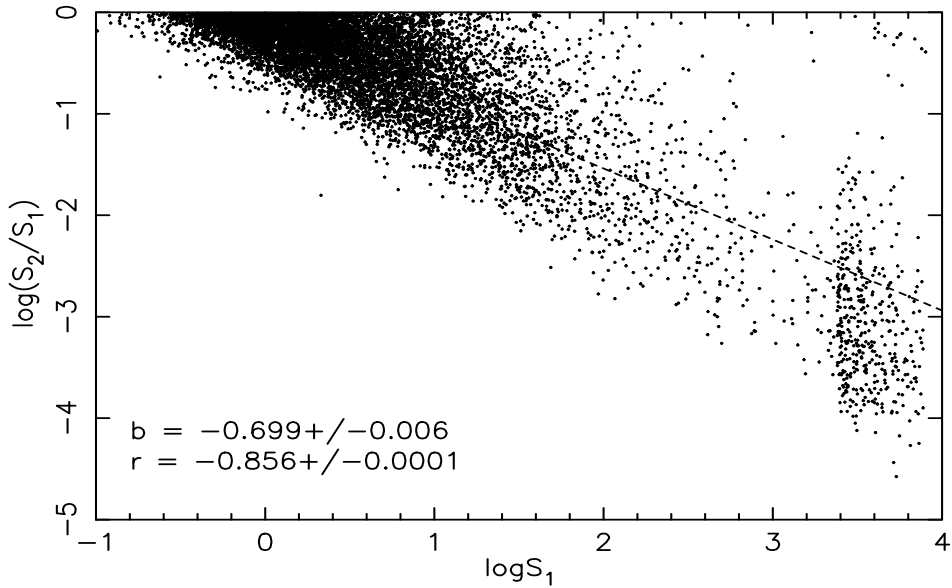}
\end{center}
\caption{\footnotesize The ratio of S2/S1 vs S1, where S1 is the peak height of 
intensity for particles from a single source and S2 is the same from the 2nd 
strongest source. The weak clustering of points observed at big values of S1 originate 
from the rare cases when the source is so close to the solar system that the latter is
inside the expanding SNR shell. Values $b$ and $r$ are given for the slope of the 
regression line and correlation coefficient respectively. }
\label{fig:fig2}
\end{figure}

In what follows we give the results of running the programmes of Erlykin and
Wolfendale, (2005) again to give the 'strengths' of both the strongest peak (S1) and
the second strongest (S2) with the result shown in Figure 2, where log(S2/S1) is plotted
 against logS1. The values of S1 were chosen to be at the energy corresponding to the
peak in the observed spectrum. The logic is that if the
contribution from S2 is small it will be unnecessary to consider the third and so on.
It is evident, and understandable, that the mean value of S2/S1 falls as S1 increases,
the point being that the larger the S1 value the less frequent it becomes; the 2nd
largest is, essentially, a 'typical source'.

Figure 3 shows the frequency distribution of S2/S1 the mean value is about 7\%. The
probability of S2/S1 being above 20\% - a value where a broadening of the 'single
source peak' might be expected to be just significant - is only $\sim$8\%.

\begin{figure}[htb]
\begin{center}
\includegraphics[height=10cm,width=12cm]{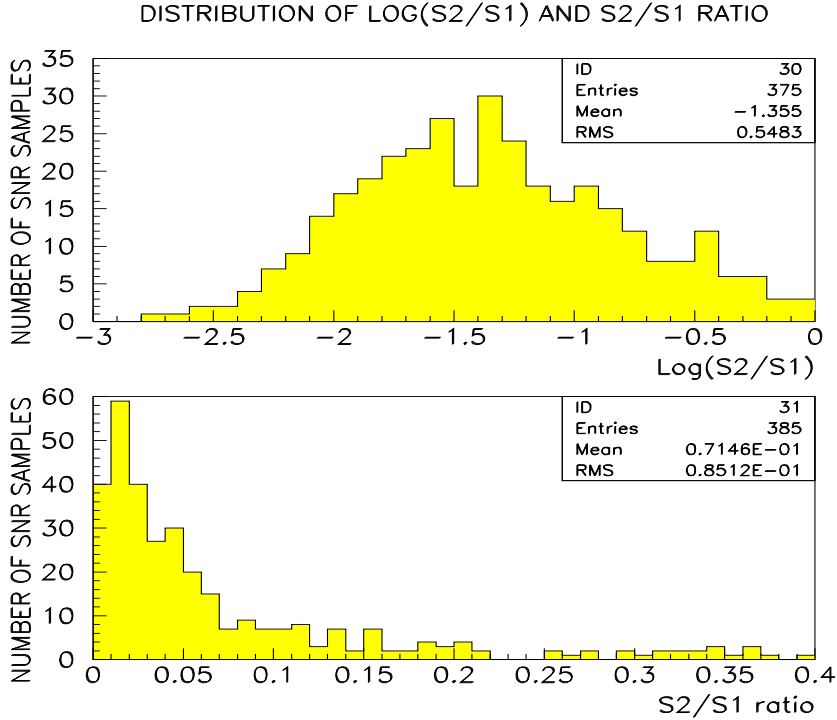}
\end{center}
\caption{\footnotesize Frequency distribution of the ratio S2/S1 of the intensity peak 
of the 2nd source, S2, to that from the single source, S1 , plotted logarithmically (a)
 and linearly (b).}
\label{fig:fig3}
\end{figure}

In an independent approach an analysis has been made of the 24 nearest likely sources
identified by Sveshnikova et al. (2011). Of the SNR, only 3 are old enough for the CR
to have any chance of arriving at Earth by now; in our model the rest are so young that
 the CR will still be trapped inside the remnant. Figure 4 shows the results. It will 
be noted that Monogem Ring ('our source') is also identified as giving the biggest 
contribution. We added 'Loop 1', although it was not selected by Sveshnikova et al, as 
the second, but its spectral shape at high energies is seen to be too steep for it to 
be relevant.
\begin{figure}[htb]
\begin{center}
\includegraphics[height=13cm,width=9cm,angle=-90]{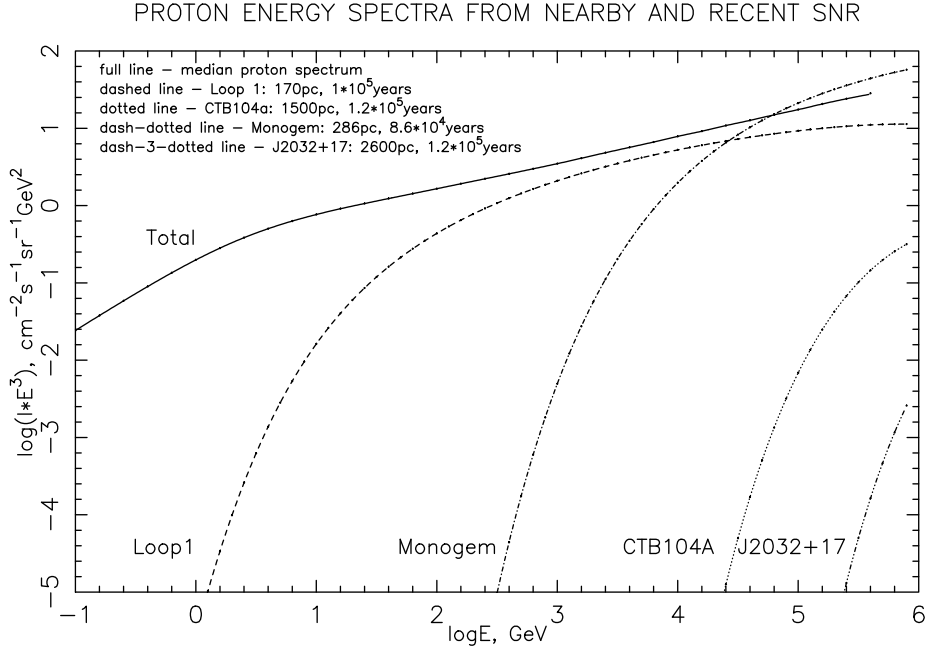}
\end{center}
\caption{\footnotesize Energy spectra from nearby and recent SNR. The SNR are chosen 
from the  set given by Sveshnikova et al.(2011) with Loop1 added. The median spectrum
of protons from 50 samples shown in Figure 5 and denoted as 'total' is given for 
comparison.}
\label{fig:fig4}
\end{figure}
\subsection{The search for `Pulsar-peaks'.}
If it is assumed that pulsars generate CR of a unique energy at each moment in their 
lives then the CR energy spectrum should show spikes and if their magnitude can be 
determined, an estimate could be made of the total pulsar contribution.

Starting with the experimental data, the precise measurements with the PAMELA 
instrument
(Adriani et al., 2011), can be considered. For protons, the intensities are given with 
statistical (and other) errors of 3\%. Inspection of the fluctuations shows none to be 
significantly in excess of this value, although there is a small excess at about 50 
GeV, which may relate to our earlier claim (Erlykin et al., 2000) of a small peak in
 the AMS data. For the range 10 to 300 GeV, beyond which the indicated errors become 
too big, we can take 3\% as the upper limit to a spike, the width being the inter-point
 separation of $\Delta logE$ = 0.05.

Although modelling of the CR spectrum above logE = 8.0 due to pulsars has been made by
us (Erlykin et al., 2001) and the frequency distribution there can be used to estimate 
from the few strongest peaks what the total would be, such an analysis has not yet been
 made in the 10-300 GeV region. All that can be said at this stage is that the total 
pulsar contribution is very unlikely to be greater than about 10\%.
\subsection{The `curvature' in the energy spectrum.}
In recent years it has become clear that the fine structure of the CR energy spectrum
exists not only in the knee region, but also at lower energies. 
Very precise observations with the PAMELA, CREAM and ATIC instruments (Adriani 
et al., 2011; Ahn et al., 2011; Panov et al., 2009) for protons, He and other nuclei 
support the contention of a concavity in the hundreds of GeV region. Specifically, the 
spectra plotted as $log(R^{3}I(R))$, where $R$ is the rigidity, have local minima at 
about 220 GV. 

Since the results of these three experiments are very close to each other for the most 
abundant proton and helium nuclei, we can analyse them together. Figure 5 
shows the actual intensities. We fitted these spectra using the formula of 
Ter-Antonyan and Haroyan, (2000). It assumes that spectra below and above the 
minima can be described by power laws with slope indices $\gamma_1$ and $\gamma_2$,
 respectively and with a smooth transition between these two slopes. The fit comprises 
5 free parameters. The obtained values of the slope indices are 
$\gamma_1^P = 2.79\pm 0.03$ and $\gamma_2^P = 2.69\pm 0.04$ for protons and 
$\gamma_1^{He} = 2.64\pm 0.03$ and $\gamma_2^{He} = 2.56\pm 0.04$ for helium nuclei, 
which is in agreement with values found in the constituent experiments of PAMELA, CREAM
 and ATIC.  

Before continuing it is necessary to consider Figure 5 in some detail. There is the 
standard problem of the undoubted presence of systematic errors in some or all of the
sets of data. For example, the CREAM and ATIC intensities are inconsistent for energies
 above the ankle. In the event this is not too important in that we need only the value
 of $\gamma_2$ here, for which the mean line seems appropriate (~ie we assume that both
 are equally accurate~). 

More important is the presence of the ankle. It will be noted that the two sets of data
(~PAMELA and ATIC~) which cover the ankle energy range have have similar slope changes,
 although that for PAMELA is sharper. Presumably, the systematic errors have little 
effect on the change of slope at the ankle, rather they give a rotation in the whole 
spectrum from each set of observations.

The joint analysis of these spectra confirms the conclusions made 
separately by these collaborations that \\
(i) the spectrum of helium nuclei is flatter than the spectrum of protons; \\
(ii) both proton and helium spectra have a concavity at an energy of hundreds of  
GeV and change their slope by about $\Delta \gamma = -0.10\pm 0.05$. \\
The negative sharpness $S$ of the concavity is very high: for both spectra it is  
$S\approx -6$, so that it is a real 'ankle' in both cases.

Interestingly an extrapolation of the best fit proton and helium spectra found by the
joint analysis of PAMELA, CREAM and ATIC data up to the knee energy of 3-4 PeV, shows 
that helium becomes the dominant component in the all-particle spectrum, with the 
fraction reaching a value of about 2/3.
\begin{figure}[htb]
\begin{center}
\includegraphics[height=13cm,width=9cm,angle=-90]{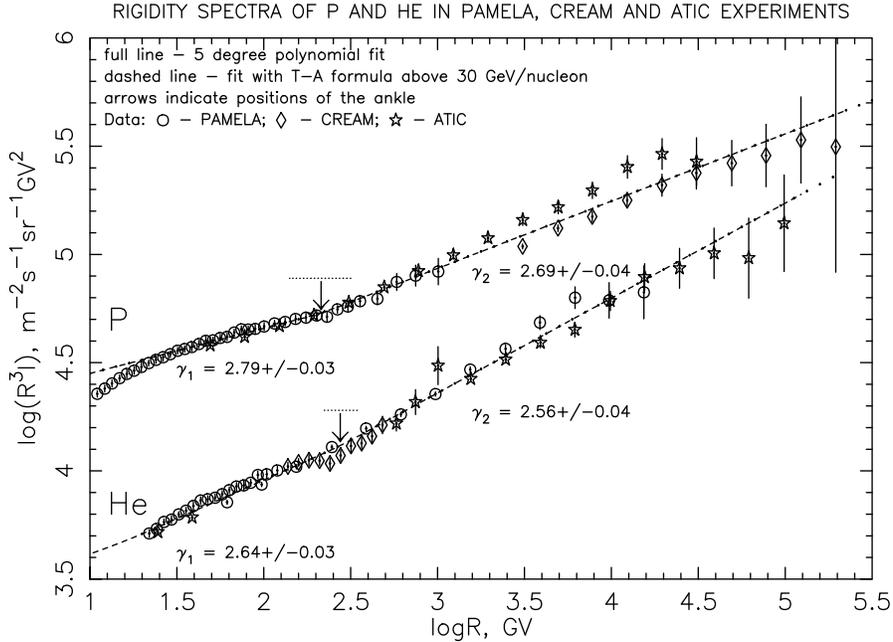}
\end{center}
\caption{\footnotesize Rigidity spectra of primary protons and helium nuclei from 
PAMELA (open circles), CREAM-I (diamonds) and ATIC-2 (stars) experiments. Arrows 
indicate the positions of the ankles, as determined by the best fitting of these 
spectra with the Ter-Antonyan and Haroyan's formula above 30 GeV/nucleon, shown by 
dashed lines. Horizontal dotted lines at the upper end of the arrows indicate the 
uncertainty interval of the ankle positions. Values of $\gamma_1$ and $\gamma_2$ are 
given for the slope indices of the spectra below and above the ankle, determined by the
 same best fit.}
\label{fig:fig5}
\end{figure}

Measurements of the energy spectra of other nuclei also show a consistent concave shape
over the common range 30-1000 GeV/nucleon (virtually the same rigidity) from the CNO 
group to Iron (using the spectra summarised by Biermann et al., 2010) and those 
considered here, although the degree of concavity is variable. We have quantified the 
concavity using a sagitta for an assumed circular shape on the plot of log($E^3I$) vs 
logE, between the limits above. 
The frequency distribution of $\delta$ from our model involving randomly distributed 
SNR has been derived from the plots given in Erlykin and Wolfendale, (2005); these are 
for the energy spectra of CR protons - but are applicable for any nucleus, replacing 
energy by rigidity; the result came from 50 trials, each involving 50,000 SNR (~Figure 
6~). The concave or convex shape of simulated spectra in the 30-1000 GeV interval 
depends on the particular pattern of the non-uniform SNR distribution in space and 
time. However, in all 50 cases variations of spectral shape are very smooth with the 
maximum (negative) sharpness $S\approx -0.6$ and the distribution of $\delta$ values 
shown in Figure 7. This distribution is symmetric around $\delta = 0$ with about equal 
numbers of positive and negative sagittas. 
\begin{figure}[htb]
\begin{center}
\includegraphics[height=13cm,width=9cm,,angle=-90]{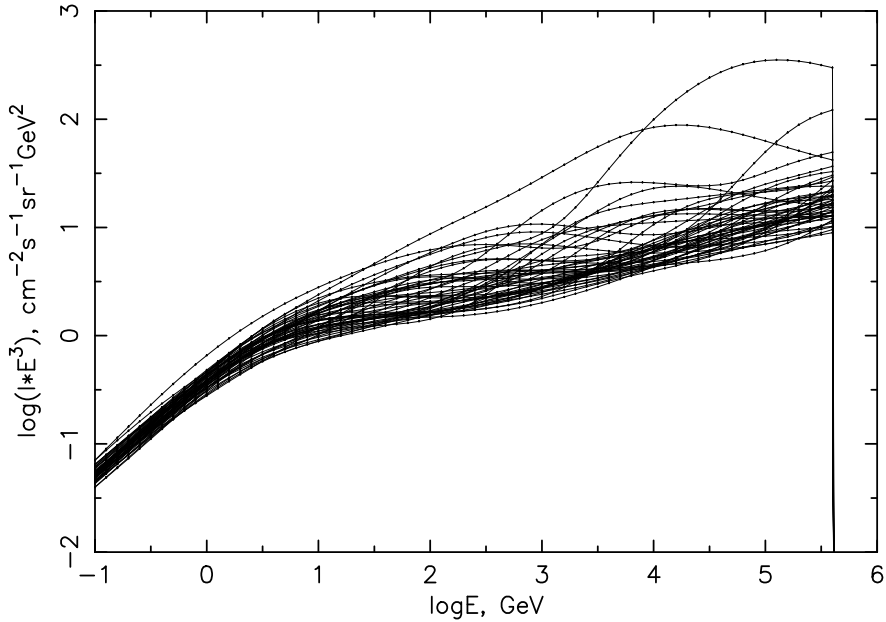}
\end{center}
\caption{\footnotesize Energy spectra for 50 samples of 50,000 randomly distributed 
SN(Erlykin and Wolfendale, 2005). The spectra would apply to any Z-value by treating 
`E' as rigidity (E/Z).}
\label{fig:fig6}
\end{figure}

The new experimental data are sufficiently precise to enable $\delta$-values to be 
determined for individual elements and there are shown in Figure 7. The experimental 
values for all nuclei including protons and helium are seen to be on the negative side.
 It is clear that the observed curvature can be hardly explained by the random 
geometric configuration of CR sources. Indeed, the probablity of such or even 
higher negative curvature for CR nuclei does not exceed 16-20\%, particularly when it 
is realised (see Figure 6) that many of the random source model spectra with large 
negative sagittas have shapes which would not fit the measured spectra at energies 
above 1000 GeV. 

The strongest argument against our random source model being the cause of actual 
sagittas is the detailed structure (and sharpness) of the ankle. In the model, the 
changes of slope are all smooth (Figure 6), but in the observations they are not as can
 be seen in Figure 5. The measured spectra for nuclei show similar sharp ankles (~eg 
the CREAM data of Ahn et al., 2010~).

Whatever the reason for the ankles, they are a further example of a fine 
structure of the CR energy spectrum and its origin, which is unlikely to be due solely 
to the non-uniformity of the SNR space-time distribution, needs a more detailed 
analysis. 
\begin{figure}[htb]
\begin{center}
\includegraphics[height=8cm, width=12cm]{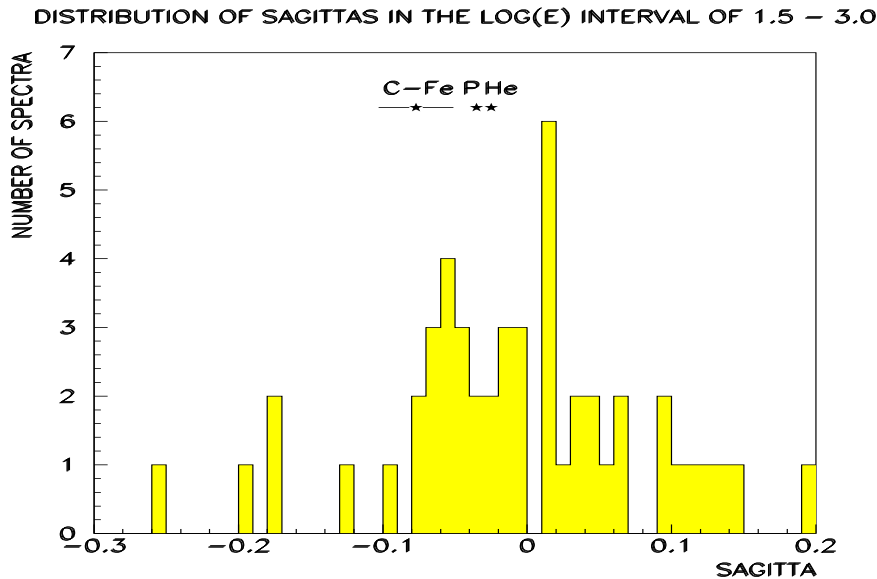}
\end{center}
\caption{\footnotesize Frequency distribution of the sagitta 
$\delta = \Delta log(E^3I(E))$ for the CR energy spectrum between 30 and 1000 
GeV/nucleon. The values derived from our analysis of data from PAMELA, CREAM and ATIC 
experiments are indicated by the stars.}
\label{fig:fig7}
\end{figure}
\section{Cosmic Ray Abundances}
There has been progress concerning the abundances of CR nuclei in the knee region as
well as at lower energies and this can be reported.

In our recent paper (Erlykin and Wolfendale, 2011b) we showed that the relative 
abundances for the single source (~Monogem Ring~) at PeV energies were essentially the 
same as those of the ambient CR at production, as measured at 1000 GeV/nucleon. This 
suggests that there is probably only a modest variation from one source (SNR) to 
another in terms of the relative abundances, although a complication is that the 
magnitude of the sagitta in the curvature of the spectra in the much lower range 
30-1000 Gev/nucleon (~see previous section~) varies from one nuclear mass to another.

Doubtless there are some (modest) differences between different sources with regard to
relative abundances just as there are differences in energetics, maximum rigidities,
etc. and these remain to be elucidated by more precise observations. Such abundance 
variations are expected from a number of causes: Interstellar Medium changes from place
 to place, as already remarked,
the possible role of pulsars in injecting energetic iron nuclei into the SNR shocks,
and variable injection of different nuclei from the pre-SN stellar winds and from the
SN ejecta.

In parenthesis, it can be remarked that the observed spectra of different elements 
could be different because the different types of SNR contributing may well have
different spatial and temporal distributions. In the limit it could be argued that 
different nuclei correspond to different spectra in Figure 6. In this case the 
difference between the proton and helium spectra in Figure 5 can be explained, and, in 
Figure 7, the average sagitta for C-Fe should be compared with a much narrower 
frequency distribution. The well known flattening of the spectra of the heavier nuclei 
above the ankle, with respect to $P$ and $He$ may find the explanation in these terms, 
too, although injection differences give an alternative explanation. 

\section{Discussion and Conclusions.}
Starting with the problem of `plural' sources in the knee region, ie the extent to
which the single source has a few accompanying contributors, we conclude that probably 
less than about 10$\%$ of the `signal' represented by the knee is due to other, 
comparatively nearby, sources. The reduction in sharpness due to this cause is thus 
expected to be very small.

Continuing with pulsars, as distinct from SNR as the progenitors of CR up to and 
including the knee, the likelihood of a significant contribution appears to be very 
small. Concerning `the search for pulsar 'peaks' at other energies which could be 
present even if pulsars are not the main sources of CR, the situation is still open. 
None has been detected so far, although, as we have pointed out (Erlykin and 
Wolfendale, 2011a) it is not impossible that the `iron peak' at $\sim$70PeV could be
due to a pulsar, rather than an SNR. It is in the region beyond the knee (where an 
examination of published spectra suggests another `bump' in the spectrum above 
logE=7.8) where pulsar peaks should be visible if, following Bednarek and
Bartosik (2005), the bulk of CR in this region come from pulsars. The argument
favouring discernable peaks about 10 PV or so is not only that pulsars of unique energy
 produce sharply peaked energy spectra (delta-functions in simple models) and that
there should only be a few such sources (~unusually short pulsar periods are
required~) but that the inevitably short Galactic trapping times also favour fine
structure, whatever the sources may be. Precise measurements are not yet available to 
substantiate this argument.

Turning to curvature in the energy spectrum just before the knee, its presence in the
total particle spectrum has been known for a long time (see, for example, Kempa et al.,
 1974) and it is due to the imminent appearance of the peak due to the
single source. There is evidence that the curvature is finite at lower energies too,
and occurs at similar rigidity for all nuclei. The magnitude of the
sagitta is greater than the average of the values found in our earlier Monte Carlo 
calculations. The PAMELA and CREAM-2 observations of fine structure at 250 GV are very 
interesting in their own right and suggest the existence of a new, and significant, 
extra CR component below 200 GV; see the work of Zatsepin and Sokolskaya, (2006).  
(~This suggestion is the subject of contemporary work~).

Approaching 10$^5$ GV rigidities the
model of Berezhko et al.(1996) predicts curvature, see, for example, Berezhko and
V\"{o}lk (2007) for a comparison of the measured proton spectrum with that predicted.
The model has maximum curvature at about 2000 GeV. At higher rigidities, say $>10^5$GV,
 curvature is observed and is expected because of the onset of nearby sources, with 
their flatter (R$^{-2}$) injection spectra - which persist in the absense of serious 
diffusive losses for particles from nearby sources.

The search for fine structure in the CR spectrum will continue to be a profitable one
and we regard the search at higher energies than the knee, coupled with very precise 
measurements below it, as the next step.

{\Large{\bf Acknowledgements}}

The authors are grateful to Professors W.Bednarek, A.R.Bell, A.Bhadra, A.G.Lyne and
G.Wright for helpful comments.

The Kohn Foundation is thanked for supporting this work.

\vspace{1cm}

{\Large{\bf References}}

\begin{enumerate}
\item Adriani, O., et al., 2011, Science, \textbf{332}, 69.
\item Ahn, H.S., et al., 2010, Astrophys. J. Lett., 714, L89.
\item Bednarek, W. and Bartosik, M., 2005, 29th ICRC (Pune) \textbf{6}, 349.
\item Bell, A.R., 2004, Mon. Not. Roy. Astron. Soc., \textbf{353}, 550.
\item Berezhko, E.G., Elshin, V.K. and Ksenofontov, L.T., 1996, J.Exp.Theor.Phys.,
\textbf{82}, 1.
\item Berezhko, E.G. and V\"{o}lk, H.J., 2007, Astrophys, J. 661, L175.
\item Bhadra, A., 2006, Astropart. Phys., \textbf{25}, 226.
\item Biermann, P.L., et al., 2010, Astrophys. J., \textbf{725}, 184.
\item Erlykin, A.D. and Wolfendale, A.W., 1997, J.Phys.G., \textbf{23}, 979.
\item Erlykin, A.D. and Wolfendale, A.W., 2001, J.Phys.G., \textbf{27}, 1005.
\item Erlykin, A.D. and Wolfendale, A.W., 2002, J.Phys.G., \textbf{28}, 359.
\item Erlykin, A.D. and Wolfendale, A.W., 2005, J.Phys.G., \textbf{31}, 1475.
\item Erlykin, A.D. and Wolfendale, A.W., 2006, J.Phys.G., \textbf{32}, 1.
\item Erlykin, A.D. and Wolfendale, A.W., 2011a, Astrophys. Space Sci. Trans, 2011, 7, 
145.
\item Erlykin, A.D. and Wolfendale, A.W., 2011b, J.Phys.G. (submitted).
\item Erlykin, A.D., Fatemi, S.J. and Wolfendale, A.W., 2000, Phys. Lett. B, 
\textbf{482}, 337.
\item Erlykin, A.D., Wibig, T. and Wolfendale, A.W., 2001, New Journ. Phys.,\textbf{3},
18.1.
\item Giller, M., and Lipski, M., 2002, J.Phys.G., \textbf{28}, 1275.
\item Kempa, J., Wdowczyk, J. and Wolfendale, A.W., 1974, J.Phys.A,\textbf{7}, 1213.
\item Kuzmichev, L.A. et al., 2011, http://tunka.sinp.msu.ru/en/presentation/Kuzmichev.pdf
\item Martirosov, R.M. et al., 2011, http://tunka.sinp.msu.ru/en/presentation/Martirosov.pdf
\item Panov, A.D., at al., 2009, Bull. Rus. Acad. Sci.: Physics, \textbf{73}, 564.
\item Sveshnikova, L.G., Ptuskin, V.S. and Strelnikova, O.N., 2011, Bull. Rus. Acad. Sci.: Physics, \textbf{75}, 334.
\item Ter-Antonyan, S.V. and Haroyan, L.S., 2000, arxiv: hep-ex 0003006.
\item Zatsepin, V.I. and Sokolskaya, N.V., 2006, Astron. and Astrophys., \textbf{458}, 1
\end{enumerate}

\end{document}